\documentclass[prd,aps,floats,twocolumn,superscriptaddress]{revtex4}

\usepackage[dvips]{graphicx}
\usepackage{amssymb}
\usepackage{amsmath}

\usepackage{color}

\begin{document}

\title{Tidal disruption rate of stars by spinning supermassive black holes}

\author{Michael Kesden} \email{mhk10@nyu.edu}

\affiliation{Center for Cosmology and Particle Physics, Department of Physics,
New York University, New York, New York 10003}

\date{August 2011}
                            
\begin{abstract}
A supermassive black hole can disrupt a star when its tidal field exceeds
the star's self-gravity, and can directly capture stars that cross its event
horizon.  For black holes with mass $M \gtrsim 10^7 M_\odot$, tidal
disruption of main-sequence stars occurs close enough to the event horizon
that a Newtonian treatment of the tidal field is no longer valid.  The fraction of
stars that are directly captured is also no longer negligible.  We calculate
generically oriented stellar orbits in the Kerr metric, and evaluate the
relativistic tidal tensor at the pericenter for those stars not directly captured by
the black hole.  We combine this relativistic analysis with previous calculations
of how these orbits are populated to determine tidal-disruption rates for
spinning black holes.  We find, consistent with previous results, that black-hole
spin increases the upper limit on the mass of a black hole capable of tidally
disrupting solarlike stars to $\sim 7 \times 10^8 M_\odot$.  More quantitatively,
we find that direct stellar capture reduces tidal-disruption rates by a factor
$\sim 2/3~(1/10)$ at $M \simeq 10^7 (10^8) M_\odot$.  The strong dependence
of tidal-disruption rates on black-hole spin for $M \gtrsim 10^8 M_\odot$ implies
that future surveys like the Large Synoptic Survey Telescope  that discover
thousands of tidal-disruption events can constrain supermassive black-hole spin demographics.
\end{abstract}

\maketitle

\section{Introduction} \label{S:intro}

In 1943, active galactic nuclei (AGN) were discovered with emission
lines Doppler broadened to widths $\gtrsim 1,000$ km/s
\cite{Seyfert:1943}.  Twenty years later, theorists proposed that these
AGN were powered by accretion onto compact objects of masses $10^5 -
10^8 M_\odot$ \cite{Hoyle:1963}.  Such massive objects cannot support
themselves against gravitational collapse into supermassive black
holes (SBHs) \cite{LyndenBell:1969yx}.  SBH masses are tightly
correlated with the luminosity \cite{Kormendy:1995er}, mass
\cite{Magorrian:1997hw}, and velocity dispersion \cite{Gebhardt:2000fk}
of the spheroidal component of their host galaxies.

SBHs primarily grow by accreting gas driven into galactic centers by
tidal torques during major mergers \cite{Toomre:1972vt,Barnes:1991zz}.
However, SBHs can also grow by directly capturing stars that cross
their event horizons or by accreting debris from stars passing close
enough to be tidally disrupted \cite{Frank:1976uy}.  Such
tidal-disruption events (TDEs) could also power bright flares of
radiation as the stellar debris is shock heated and subsequently
accreted \cite{Rees:1988bf,Strubbe:2010kq}.  Several potential TDEs
have been found in x-ray \cite{Bade:1996}, UV \cite{Gezari:2006fe},
and optical \cite{vanVelzen:2010jp} surveys, and tidal debris may also fuel
recent blazar activity seen by the Swift satellite
\cite{Burrows:2011dn,Levan:2011yr,Bloom:2011xk,Cenko:2011ys}.
The handful of TDEs found in current surveys implies that thousands more
may be found each year in future surveys by the Large Synoptic Survey
Telescope \cite{vanVelzen:2010jp,:2009pq}.  A detailed
comparison between predicted and observed TDE rates will provide
important constraints on SBHs and the galactic centers in which they
reside.

Frank and Rees \cite{Frank:1976uy} were among the first to estimate
TDE rates.  They introduced the concept of a "loss cone" in the stellar
phase space that would be depopulated by tidal disruption within a
dynamical time $t_{\rm dyn}$.  Stars within the loss cone have velocities
lying in a cone about the radial direction.  TDE rates are set by the rate at
which stellar diffusion from other portions of phase space refills this loss cone.
Frank and Rees evaluated stellar fluxes into the loss cone at a critical
radius $r_{\rm crit}$ at which stellar diffusion operating on a reference
time scale $t_R$ \cite{Spitzer:1958} could refill the loss cone within
$t_{\rm dyn}$.  Cohn and Kulsrud \cite{Cohn:1978} provided a more
sophisticated treatment of stellar diffusion into the loss cone by numerically
integrating the Fokker-Planck equation in energy-angular momentum space.
More recently, Wang and Merritt \cite{Wang:2003ny} have revised predicted
TDE rates using more realistic galactic density profiles and the observed
relation between SBH mass and host-galaxy velocity dispersion
\cite{Gebhardt:2000fk}.

These analyses focused on smaller SBHs for which a Newtonian treatment
of tidal forces is valid and the number of directly captured stars is negligible
compared to the number that are tidally disrupted.  Manasse and Misner 
\cite{Manasse:1963} introduced Fermi normal coordinates that are ideal for
a relativistic treatment of the tidal tensor, and calculated this tidal tensor for
radial geodesics of the Schwarzschild metric for nonspinning SBHs.  Marck
\cite{Marck:1983} generalized this calculation to generically oriented timelike
geodesics of the Kerr metric \cite{Kerr:1963ud} for spinning SBHs.
Beloborodov {\it et al.} \cite{Beloborodov:1992} used this tidal tensor to
calculate the relativistic cross sections for tidal disruption for a range of initial
orbital inclinations with respect to the SBH spin.  Ivanov and Chernyakova
\cite{Ivanov:2005se} used a numerically fast Lagrangian model of a tidally
disrupted star to investigate how stellar hydrodynamics affects these relativistic
cross sections.  In this paper, we combine a similar relativistic treatment of
tidal disruption and direct capture with existing calculations of loss-cone physics
to derive improved predictions of TDE rates for massive and highly spinning SBHs.

The first step in calculating TDE rates is to establish criteria for
determining when a star is tidally disrupted.  Most tidally disrupted
stars are initially on highly eccentric or unbound orbits
characterized by the distance $r$ of their pericenters from the SBH.
An order-of-magnitude estimate of the maximum value of $r$ for which
tidal disruption occurs can be obtained by equating the differential
acceleration $GMR_\ast/r^3$ experienced by a star of mass $m_\ast$ and
radius $R_\ast$ in the tidal field of a black hole of mass $M$ to the
star's self-gravity $Gm_\ast/R_{\ast}^{2}$.  This implies that a star
will be tidally disrupted when $r < r_{\rm TD} \simeq (M/m_\ast)^{1/3}
R_\ast$.  A star will be directly captured by the SBH when $r$ is less
than the radius of the event horizon, which for a nonspinning SBH
is equal to the Schwarzschild radius $r_{\rm S} = 2GM/c^2$.
Since $r_{\rm TD} \propto M^{1/3}$ while $r_{\rm S} \propto M$, the
ratio of tidally disrupted to directly captured stars will decrease
with increasing SBH mass.  Equating these two distances, we find that
a SBH with mass $M$ greater than
\begin{eqnarray}
M_{\rm max} &\simeq& \frac{c^3}{m_{\ast}^{1/2}}
\left( \frac{R_\ast}{2G} \right)^{3/2}
\nonumber \\
\label{E:Mmax}
&=& 1.1 \times 10^8 M_\odot \left( \frac{m_\ast}{M_\odot} \right)^{-1/2}
\left( \frac{R_\ast}{R_\odot} \right)^{3/2}
\end{eqnarray}
should directly capture stars instead of tidally disrupting them.

Our estimate of $r_{\rm TD}$ assumed that the gravitational field of
the SBH was that of a Newtonian point particle, which should only be
valid for $r_{\rm TD} \gg r_{\rm S}$.  One should be {\it very}
suspicious of using this estimate at the event horizon, as we did when
deriving $M_{\rm max}$ above.  In a proper general-relativistic
treatment, the spacetime of a spinning SBH is described by the Kerr
metric \cite{Kerr:1963ud}.  The Kerr metric is a two-parameter family of
solutions to Einstein's equation fully specified by the SBH mass $M$
and dimensionless spin $a/M < 1$.  Theoretical estimates of SBH spins depend
sensitively on the extent to which SBHs grow by accretion or mergers.
SBHs accreting from a standard thin disk \cite{Shakura:1972te} can
attain a limiting spin $a/M \simeq 0.998$ \cite{Thorne:1974ve} after
increasing their mass by a factor $\sim \sqrt{6}$ \cite{Bardeen:1970}.
The spins of SBHs formed in mergers vary greatly depending on
whether the spins of the initial binary black holes become aligned
with their orbital angular momentum prior to merger
\cite{Berti:2008af}.  SBH spins can be inferred from observations of
iron K$\alpha$ lines in AGN x-ray spectra \cite{Reynolds:1998ie}.
Large spins have been measured, such as $a/M =
0.989_{-0.002}^{+0.009}$ in the Seyfert 1.2 galaxy MCG-06-30-15
\cite{Brenneman:2006hw}, although reliable estimates are only
available for a small number of systems.

Given the large sample of observed TDEs expected in the near future
and the wide range of predicted SBH spins, it is important to
determine the extent to which TDE rates depend on SBH spin.  This is
the primary goal of this paper.  The greater the spin dependence, the more
tightly observed TDEs will be able to constrain the distribution of SBH
spins.  In Sec.~\ref{S:geo}, we review how the tidal field is
calculated along timelike geodesics of the Kerr metric.  In
Sec.~\ref{S:sims}, we describe the Monte Carlo simulations we
performed to determine which stellar orbits lead to tidal disruption.
We then use these simulations to calculate expected TDE rates in
Sec.~\ref{S:rates}.  The implications of our findings are discussed in
Sec.~\ref{S:disc}.

\section{Tidal fields along Kerr geodesics} \label{S:geo}

In Boyer-Lindquist coordinates \cite{Boyer:1966qh} and units where $G
= c = 1$, the Kerr metric takes the form
\begin{eqnarray} \label{E:met}
ds^2 &=& -\left( 1 - \frac{2Mr}{\Sigma} \right) dt^2
- \frac{4Mar \sin^2 \theta}{\Sigma} dt d\phi
+ \frac{\Sigma}{\Delta} dr^2
\nonumber \\ 
&&  + \Sigma d\theta^2 + \left( r^2 + a^2 +
\frac{2Ma^2r \sin^2 \theta}{\Sigma} \right) \sin^2 \theta d\phi^2
\end{eqnarray}
where $\Sigma \equiv r^2 + a^2 \cos^2 \theta$ and $\Delta \equiv r^2 -
2Mr + a^2$.  This metric is both stationary (independent of $t$) and
axisymmetric (independent of $\phi$).  Massive test particles travel
on timelike geodesics of the Kerr metric.  Individual stars have
masses $m_\ast \sim M_\odot$ much less than those of SBHs ($10^6
M_\odot \lesssim M \lesssim 10^{10} M_\odot$), and radii $R_\ast \sim
R_\odot \simeq 7 \times 10^{10}$ cm less than the Schwarzschild radius
\begin{equation} \label{E:RS}
r_{\rm S} = \frac{2GM}{c^2} = 2.95 \times 10^{11}~{\rm cm} \left(
\frac{M}{10^6 M_\odot} \right)~.
\end{equation}
We can therefore consider them to be test particles for the purpose of
determining their orbits.  The position $(r, \theta, \phi)$ of a star
as a function of proper time $\tau$ evolves according to the equations
\cite{Carter:1968rr}
\begin{subequations} \label{E:EOM}
\begin{eqnarray}
\Sigma^2 \left( \frac{dr}{d\tau} \right)^2 &=& [E(r^2 + a^2) - aL_z]^2
\nonumber \\
\label{E:drdt}
&& -\Delta[r^2 + (L_z - aE)^2 + Q] \\
\label{E:dthetadt}
\Sigma^2 \left( \frac{d\theta}{d\tau} \right)^2 &=&
Q - L_{z}^2 \cot^2 \theta - a^2 (1-E^2) \cos^2 \theta \\
\label{E:dphidt}
\Sigma \left( \frac{d\phi}{d\tau} \right) &=&
L_z \csc^2 \theta + \frac{2MarE}{\Delta} -\frac{a^2 L_z}{\Delta}~,
\end{eqnarray}
\end{subequations}
where the specific energy $E$, angular momentum $L_z$, and Carter
constant $Q$ are conserved along geodesics.

Although Boyer-Lindquist coordinates reduce to flat-space spherical
coordinates in the limit $r \to \infty$, the nonzero off-diagonal
elements of the Kerr metric (\ref{E:met}) imply that these coordinate
vectors do not constitute an orthogonal tetrad at finite $r$.  The
gravitational-field gradients experienced by freely falling observers
are more conveniently expressed by projecting them onto an orthonormal
tetrad $\boldsymbol{\lambda}_\mu$ like that provided by Fermi normal
coordinates \cite{Manasse:1963}.  This coordinate system can be used
to specify points in the neighborhood of a central timelike geodesic,
such as that traversed by a star orbiting a Kerr SBH.  The timelike
member of this tetrad $\boldsymbol{\lambda}_0$ is the tangent vector
along the central geodesic, while the spacelike vectors
$\boldsymbol{\lambda}_i$ ($i = 1, 2, 3$) span the plane in the tangent
space othogonal to $\boldsymbol{\lambda}_0$.  The point ($\tau, x^i$)
in Fermi normal coordinates is reached by starting at the location of
the star at proper time $\tau$ and moving a proper distance $R =
\sqrt{\sum_i (x^i)^2}$ along the spacelike geodesic whose tangent
vector is $\sum_i x^i \boldsymbol{\lambda}_i$.

In Fermi normal coordinates, the time-time component of the metric can
be Taylor expanded as
\begin{equation} \label{E:ttFNC}
g_{\mu\nu} \lambda_{0}^\mu \lambda_{0}^\nu = -1 - R_{0i0j} x^i x^j + ...~,
\end{equation}
where $R_{\alpha\beta\gamma\delta}$ is the Riemann curvature tensor
projected onto the orthonormal tetrad $\boldsymbol{\lambda}_\mu$.
This implies that the tidal potential $\Phi_{\rm tidal}$ experienced
by a star is
\begin{equation} \label{E:tidpot}
\Phi_{\rm tidal} = -\frac{1}{2}(g_{\mu\nu} \lambda_{0}^\mu \lambda_{0}^\nu
+ 1) = \frac{1}{2} C_{ij} x^i x^j + ...~,
\end{equation}
where $C_{ij} \equiv R_{0i0j}$ is the tidal tensor.  Although the
higher-order corrections to the tidal potential denoted by the
ellipsis can sometimes be significant \cite{Ishii:2005xq}, in this
paper we consider only the term quadratic in $x^i$.  The tidal tensor
$C_{ij}$ is a symmetric, traceless $3 \times 3$ matrix whose
eigenvectors denote the principal axes along which the star is
stretched or squeezed, and whose eigenvalues denote the extent of this
stretching and squeezing.

The problem of calculating the tidal potential $\Phi_{\rm tidal}$ thus
reduces to choosing an appropriate orthonormal tetrad
$\boldsymbol{\lambda}_\mu$ for generic Kerr geodesics and projecting
the Riemann tensor onto this tetrad.  This has already been
accomplished for us by Marck \cite{Marck:1983}, who found
\begin{subequations} \label{E:TT}
\begin{eqnarray}
C_{11} &=& \left(1 - 3\frac{ST(r^2 - a^2 \cos^2 \theta)}{K\Sigma^2}
\cos^2 \Psi \right) I_1 \nonumber \\
\label{E:C11}
&& + 6ar\cos \theta \frac{ST}{K\Sigma^2} \cos^2 \Psi I_2 \\
C_{12} &=& [-ar\cos \theta (S+T)I_1 \nonumber \\
\label{E:C12}
&& + (a^2 \cos^2 \theta S - r^2 T) I_2] 3\frac{\sqrt{ST}}{K\Sigma^2}\cos \Psi \\
C_{13} &=& [(a^2 \cos^2 \theta - r^2)I_1 \nonumber \\
\label{E:C13}
&& + 2ar \cos \theta I_2] 3\frac{ST}{K\Sigma^2}\cos \Psi \sin \Psi \\
C_{22} &=& \left(1 + 3\frac{r^2T^2 - a^2\cos^2 \theta S^2}{K\Sigma^2}
\right) I_1 \nonumber \\
\label{E:C22}
&& - 6ar\cos \theta \frac{ST}{K\Sigma^2} I_2 \\
C_{23} &=& [-ar\cos \theta (S+T)I_1 \nonumber \\
\label{E:C23}
&& + (a^2 \cos^2 \theta S - r^2 T) I_2] 3\frac{\sqrt{ST}}{K\Sigma^2}\sin \Psi \\
C_{33} &=& \left(1 - 3\frac{ST(r^2 - a^2 \cos^2 \theta)}{K\Sigma^2}
\sin^2 \Psi \right) I_1 \nonumber \\
\label{E:C33}
&& + 6ar\cos \theta \frac{ST}{K\Sigma^2} \sin^2 \Psi I_2~,
\end{eqnarray}
\end{subequations}
where
\begin{subequations}
\begin{eqnarray}
\label{E:KTT}
K &\equiv& (L_z - aE)^2 + Q \\
\label{E:STT}
S &\equiv& r^2 + K \\
\label{E:TTT}
T &\equiv& K - a^2 + \cos^2 \theta \\
\label{E:I1}
I_1 &\equiv& \frac{Mr}{\Sigma^3} (r^2 - 3a^2 \cos^2 \theta) \\
\label{E:I2}
I_2 &\equiv& \frac{Ma\cos \theta}{\Sigma^3} (3r^2 - a^2 \cos^2 \theta)~.
\end{eqnarray}
\end{subequations}
The angle $\Psi$ evolves along the geodesic to ensure that
$\boldsymbol{\lambda}_1$ and $\boldsymbol{\lambda}_3$ are parallel
propagated.

The fully general tidal tensor of Eq.~(\ref{E:TT}) is intimidating,
but we can gain insight by considering the tidal tensor for equatorial
geodesics ($\theta = \pi/2, Q = 0$) whose nonzero elements are
\begin{subequations} \label{E:TTeq}
\begin{eqnarray}
\label{E:C11eq}
C_{11} &=& \left(1 - 3\frac{r^2 + K}{r^2} \cos^2 \Psi \right) \frac{M}{r^3} \\
\label{E:C13eq}
C_{13} &=& -3\frac{r^2 + K}{r^5} M \cos \Psi \sin \Psi \\
\label{E:C22eq}
C_{22} &=& \left(1 + 3\frac{K}{r^2} \right) \frac{M}{r^3} \\
\label{E:C33eq}
C_{33} &=& \left(1 - 3\frac{r^2 + K}{r^2} \sin^2 \Psi \right) \frac{M}{r^3}~.
\end{eqnarray}
\end{subequations}
The eigenvalues of this tensor are $M/r^3$, $(1+3K/r^2)M/r^3$, and
$-2(1+3K/2r^2)M/r^3$.  Since the tidal force is
\begin{equation} \label{E:Ftid}
F_i = -\nabla_i \Phi_{\rm tidal} = -C_{ij} x^j~,
\end{equation}
the positive eigenvalues correspond to directions in which the star is
squeezed while the negative eigenvalues correspond to the direction in
which it is streched.  In the Newtonian limit $K/r^2 \to 0$, the
eigenvalues reduce to $-2M/r^3$ and the doubly degenerate eigenvalue
$M/r^3$.  This degeneracy reflects the restoration of symmetry between
the $\theta$ and $\phi$ directions at large $r$, where the effects of
the SBH's spin are negligible.  Stretching occurs in the radial
direction corresponding to the eigenvalue $-2M/r^3$.  Note that despite one's
possible intuition to the contrary, the tidal force remains finite at both the
innermost stable circular orbit and even the event horizon itself.

To determine whether a star on a given orbit is tidally disrupted, we
check at the pericenter of that orbit whether the outward tidal force in
the direction corresponding to the negative eigenvalue of the tidal
tensor exceeds the inwards Newtonian self-gravity of the star.  We
assume that the tidal field is maximized at the pericenter as in the
Newtonian limit.  If $\beta_-$ denotes the numerical value of this
eigenvalue, tidal disruption occurs if
\begin{equation} \label{E:TDcon}
r < r_{\rm TD} = \left[ \left( \frac{|\beta_-|}{M/r^3} \right)
\left( \frac{M}{m_\ast} \right) \right]^{1/3} R_\ast~.
\end{equation}
In the Newtonian limit $\beta_- = -2M/r^3$ discussed above, this
condition is equivalent to the more familiar expression
\begin{equation} \label{E:TDconNR}
r < r_{\rm TD} = \left( \frac{2M}{m_\ast} \right)^{1/3} R_\ast~.
\end{equation}

Although our condition (\ref{E:TDcon}) for tidal disruption is only
approximate, we expect it to be conservative for several reasons.  It
neglects that the tidal force has already raised bulges on the star's
surface before the star reaches the pericenter, so the stellar radius
$R_\ast$ appearing in Eq.~(\ref{E:TDcon}) should exceed its value in
hydrostatic equilibrium far from the SBH.  It also assumes that the
star is nonrotating, while in reality the torques exerted on the
tidally distorted star will cause it to partially corotate with its
orbit.  These torques are likely to be complicated for a generic
nonequatorial Kerr geodesic, but we can gain some insight by again
considering the Newtonian limit.  Stars rotating with angular velocity
$\Omega$ will be disrupted at a radius
\begin{equation} \label{E:TDconrot}
r < r_{\rm TD}(\Omega) = \left[ \left( 2 + \frac{\Omega^2 r^3}{GM}
\right) \left( \frac{M}{m_\ast} \right) \right]^{1/3} R_\ast
\end{equation}
in this limit.  For corotating stars on circular orbits ($\Omega^2 =
GM/r^3$), the first factor in parentheses on the right-hand side of
Eq.~(\ref{E:TDconrot}) equals 3 as in the definition of the radius of
the Hill's sphere \cite{Murray}.  For a star corotating at the pericenter
of a parabolic orbit like that expected for a star approaching a SBH
from a large distance, this factor equals 4.  Our assumption that the
star is nonrotating is conservative because the condition
(\ref{E:TDconrot}) is most restrictive for $\Omega = 0$, although
$r_{\rm TD}(\Omega)$ only varies by the modest factor $2^{1/3}$.

Our criterion (\ref{E:TDcon}) might overestimate the rate at which
stars are fully disrupted since they might lose their outer layers
while maintaining their dense cores.  In the Newtonian limit, Phinney
\cite{Phinney} showed that stars will not be fully disrupted until
\begin{equation}
r < r_{\rm TD} = \left( \frac{k}{f} \right)^{1/6}
\left( \frac{M}{m_\ast} \right)^{1/3} R_\ast~,
\end{equation}
where $k$ is the constant of apsidal motion and $fGm_{\ast}^2/R_\ast$
is the star's binding energy.  The factor $k/f = 0.3 (0.02)$ for stars
with convective (radiative) atmospheres, but the exponent of 1/6
ensures that $r_{\rm TD}$ is only weakly dependent on this factor.  We
ignore this factor and keep our criterion (\ref{E:TDcon}) for the
remainder of this paper, but the Monte Carlo simulations described in
the next section could easily be evaluated with a new criterion that
incorporates this factor or a different choice of stellar properties
than $m_\ast = M_\odot$, $R_\ast = R_\odot$.

\begin{figure}[t!]
\begin{center}
\includegraphics[width=3.5in]{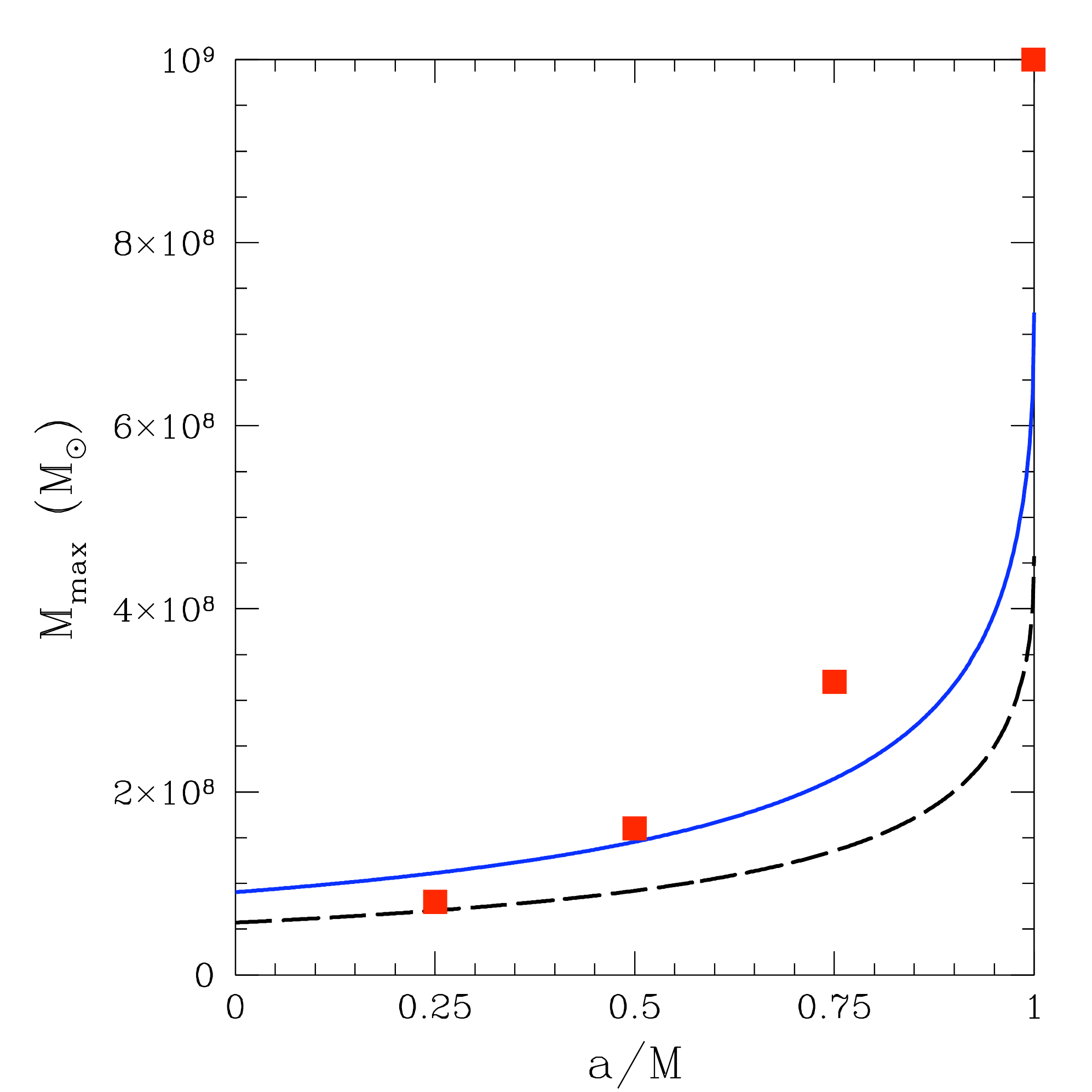}
\end{center}
\caption{The mass $M_{\rm max}$ of the heaviest SBH capable of
disrupting a star of solar mass and radius as a function of SBH spin
$a/M$.  The red squares show the values listed in Table 2 of
\cite{Ivanov:2005se} derived using a simple hydrodynamical model.
The solid blue curve shows our prediction according to the
relativistic criterion (\ref{E:TDcon}), while the dashed black curve
shows the Newtonian prediction (\ref{E:TDconNR}).}
\label{F:Mmax}
\end{figure}

A detailed study of the fraction of mass loss as a function of SBH and
orbital parameters is beyond the scope of this paper, but such a study
for selected orbital inclinations has been performed by Ivanov and
Chernyakova \cite{Ivanov:2005se}.  They recognized that for a given
SBH mass and spin, stars on prograde, equatorial, marginally bound
orbits \cite{Bardeen:1972fi} are the most likely to be tidally disrupted.
Using a simple but computationally inexpensive
hydrodynamical model, they calculated the mass $M_{\rm max}$ of the
heaviest SBH capable of tidally disrupting stars without directly capturing
them.  As in the Newtonian prediction of Eq.~(\ref{E:Mmax}), $M_{\rm max}
\propto m_{\ast}^{-1/2} R_{\ast}^{3/2}$.  In Fig.~\ref{F:Mmax}, we compare
their predictions (red squares) to our own using the relativistic criterion
(\ref{E:TDcon}) (solid blue curve) and Newtonian criterion (\ref{E:TDconNR})
(dashed black curve) for stellar mass $m_\ast = M_\odot$ and radius
$R_\ast = R_\odot$.  We see that the relativistic correction to the Newtonian
prediction is significant, and that our simple criterion (\ref{E:TDcon}) does a
reasonable job given the $\sim 50\%$ uncertainty in the simulations
\cite{Ivanov:2005se}.

We see that in the maximally spinning limit ($a/M \to 1$), a SBH as
massive as $\sim 10^9 M_\odot$ is capable of tidally disrupting
main-sequence stars.   This prediction is consistent with earlier
simulations \cite{Sponholz,Kobayashi:2004py} that demonstrated this
possibility.  The above scaling of $M_{\rm max}$ with stellar mass and
radius suggests that a white dwarf with $m_\ast \simeq M_\odot, R_\ast \simeq
0.01 R_\odot$ could be tidally disrupted by a maximally spinning SBH as
massive as $10^6 M_\odot$.  This conclusion helps alleviate tension
between the small SBH mass required for the interpretation of Swift J1644+57
as a white-dwarf tidal disruption \cite{Krolik:2011fb} and the larger value of $M$
inferred from the relation between SBH mass and host-galaxy velocity
dispersion \cite{Gebhardt:2000fk}.

\section{Monte Carlo simulations} \label{S:sims}

Unlike the Newtonian two-body problem, there is no general analytic
solution to the relativistic equations of motion (\ref{E:EOM}).  We
must integrate these equations explicitly for every orbit we consider.
Stars that will eventually be tidally disrupted are scattered onto
their final orbits at radii $r \gg r_{\rm TD}$.  These orbits may or
may not be gravitationally bound to the SBH, but their Newtonian
orbital energies $\sim m_\ast \sigma^2$, where $\sigma$ is a typical
velocity at $r \gg r_{\rm TD}$, are much less than the rest-mass
energy $m_\ast c^2$.  It is therefore an excellent approximation to
set the specific energy $E$ appearing in Eqs.~(\ref{E:EOM}) equal to
unity in units where $c = 1$.  The Kerr metric (\ref{E:met}) is
axisymmetric, so our results are independent of the initial value of $\phi$.  We must
perform Monte Carlo simulations with an appropriate distribution of
the remaining variables $\{ r, \theta, L_z, Q \}$ to determine what
fraction of orbits are tidally disrupted according to our relativistic
criterion (\ref{E:TDcon}), where the negative eigenvalue $\beta_-$ of
the tidal tensor $C_{ij}$ depends on all these variables.

\begin{figure}[t!]
\begin{center}
\includegraphics[width=3.5in]{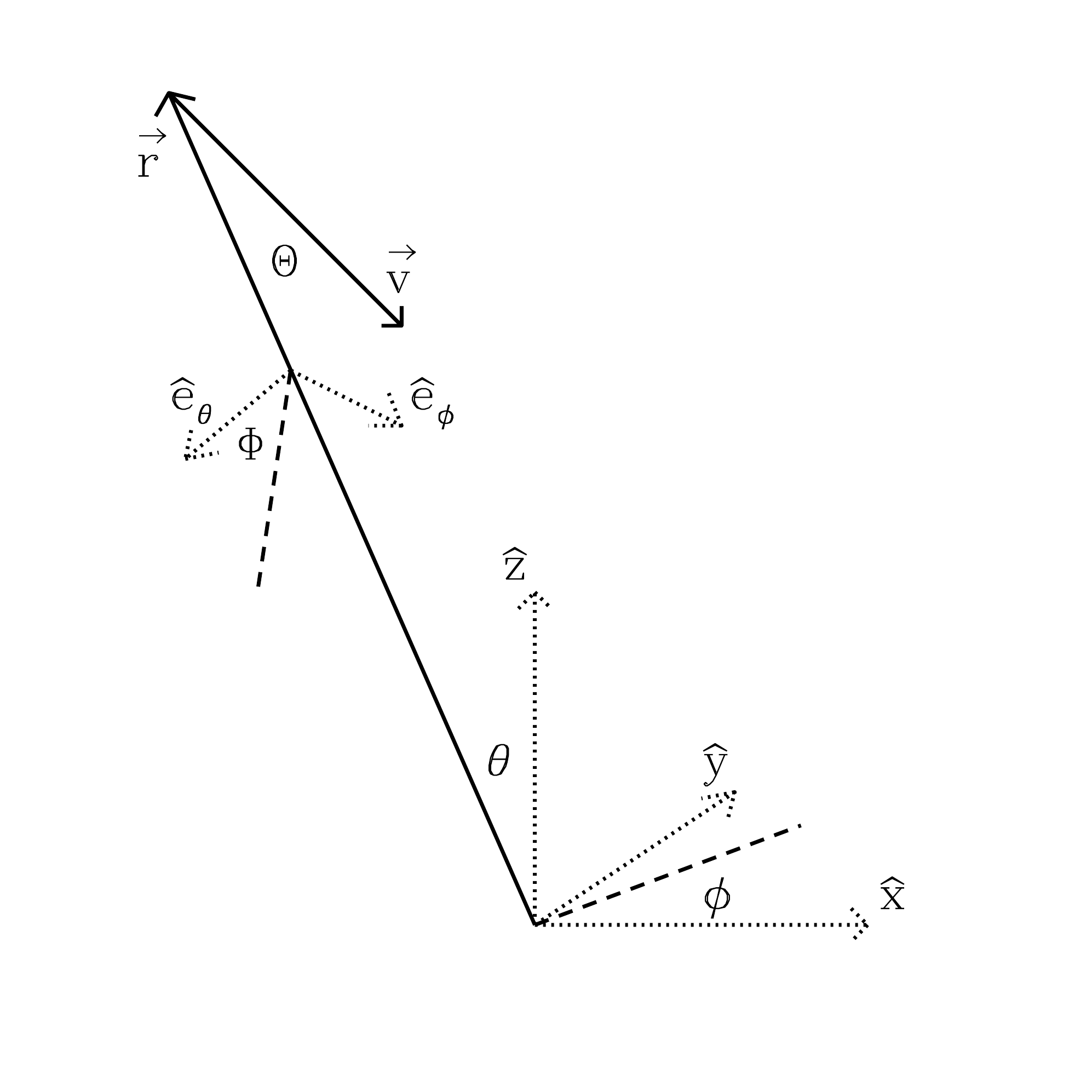}
\end{center}
\caption{Our choice of coordinates for determining the initial
conditions for integrating the equations of motion (\ref{E:EOM}) for
stellar orbits.  The SBH is located at the origin, and the star is
located at Boyer-Lindquist coordinates ($r, \theta, \phi$).  $\Theta$
is the angle between the stellar velocity $\mathbf{v}$ and the inwards
radial direction $-\mathbf{\hat{r}}$, while $\Phi$ is the angle
between the component of $\mathbf{v}$ perpendicular to
$\mathbf{\hat{r}}$ and the unit vector $\mathbf{\hat{e}}_\theta$ in
the $\theta$ direction.}
\label{F:geo}
\end{figure}

We illustrate the geometry of the problem and our choice of
coordinates in Fig.~\ref{F:geo}.  We begin integrating the equations
of motion (\ref{E:EOM}) with the star located at an initial position
($r, \theta, \phi$) in Boyer-Lindquist coordinates.  Since these
equations are independent of $\phi$ as is the tidal tensor $C_{ij}$,
we do not actually need to integrate Eq.~(\ref{E:dphidt}).  We choose
an initial radius $r = 2000 M$, where relativisic corrections are
small, and check that our results are insensitive to this choice.  In
this limit, the constants of motion are given by
\begin{subequations} \label{E:ICs}
\begin{eqnarray}
L_z &=& rv \sin \theta \sin \Theta \sin \Phi \\
\label{E:Lz}
Q &=& L_{x}^2 + L_{y}^2
\nonumber \\
&=& (rv \sin \Theta)^2 (\cos^2 \Phi + \cos^2 \theta \sin^2 \Phi)~,
\label{E:Q}
\end{eqnarray}
\end{subequations}
where $v = (2M/r)^{1/2}$ is the magnitude of the initial stellar
velocity and $\Theta$ and $\Phi$ are the angles described in
Fig.~\ref{F:geo}.  Since the stars at $r \gg r_{\rm TD}$ do not know
about the direction of the SBH spin, the stellar distribution is
axisymmetric about $\mathbf{\hat{r}}$ and there is a uniform
distribution in $\Phi$.  Although astrophysical spheroids do not
necessarily have isotropic velocity dispersions at large $r$, the
stars approaching this close to the SBH belong to the tiny
fraction of the distribution where the velocity lies in a loss cone
centered about the radial direction \cite{Frank:1976uy}.  Since there
is no reason to expect the distribution function to be varying
strongly in this small portion of phase space, there is a uniform
distribution in $-1 \leq \cos \Theta \leq 1$ as well.  However, since
the rate at which stars of velocity $\mathbf{v}$ enter the sphere of
radius $r$ is proportional to $\mathbf{v} \cdot \mathbf{\hat{r}}$, we
weight our distribution by $\cos \Theta$ during our Monte Carlo
simulations of stellar orbits.  We choose a maximum value $\Theta_{\rm
max}$ to avoid wasting computational time on orbits that do not
closely approach the SBH.

With this choice of initial conditions, we integrate the equations of
motion (\ref{E:EOM}) until the star reaches the pericenter.  We then
calculate and tabulate the negative eigenvalue $\beta_-$ of the tidal
tensor $C_{ij}$ (\ref{E:TT}).  We also tabulate which stars are
directly captured by the SBH when their orbits encounter the SBH's
event horizon.  We integrated 250,000 stellar orbits for each of
several SBH spins, with an additional 250,000 with a smaller choice of
$\Theta_{\rm max}$ to increase our sampling of the small number of
orbits that lead to tidal disruption when $M \to M_{\rm max}$.

\section{TDE rates} \label{S:rates}

Given a stellar phase-space distribution function, it is reasonably
straightforward to calculate the rate of TDEs
using the Monte Carlo simulations described in the previous
section.  In Sec.~\ref{SS:Max} below, we calculate the TDE rate as a
function of SBH mass assuming that the stars approach a Maxwellian
distribution with fixed number density and velocity dispersion far
from the SBH.  This calculation illustrates the dependence of TDE
rates on SBH spin.  However, astrophysical SBHs reside in galaxy
spheroids whose properties are tightly correlated with SBH mass
\cite{Kormendy:1995er,Magorrian:1997hw,Gebhardt:2000fk}.  In addition,
the very luminous early-type galaxies that host the most massive SBHs
have cored profiles at their centers unlike the power-law profiles
that characterize less luminous early-type galaxies and late-type
bulges \cite{Faber:1996yg}.  Predicted TDE rates are sensitive to
whether galactic centers are described by cored or power-law profiles
\cite{Wang:2003ny}.  Recent observations \cite{Buchholz:2009st}
suggest that even the nuclear star cluster at our own Galactic center,
long believed to have a cuspy profile ($\rho \propto r^{-7/4}$)
\cite{Bahcall:1976aa}, may in fact have a core of radius $r_{\rm core}
\simeq 0.5$ pc \cite{Merritt:2009mr}.  Given these uncertainties, it
is difficult to make precise estimates of astrophysical TDE rates.
Despite this, in Sec.~\ref{SS:real} we calculate the TDE rate assuming
galaxies have isothermal ($\rho \propto r^{-2}$) profiles at their
centers and host SBHs with masses correlated with their velocity
dispersions.  The results of this calculation shown in
Fig.~\ref{F:Rrate} illustrate how SBH spin affects TDE rates.

\subsection{Maxwellian distribution} \label{SS:Max}

Assume that stars far from the SBH have a Maxwellian distribution
function
\begin{equation} \label{E:maxDF}
f(\mathbf{r},\mathbf{v}) = \frac{n}{(2\pi \sigma^2)^{3/2}}
e^{-\mathbf{v}^2/2\sigma^2}
\end{equation}
with number density $n$ and velocity dispersion $\sigma$.  The
differential rate at which stars with Newtonian specific energy $E_N$
and angular momentum $L_N$ enter a sphere of radius $r$ is given by
\begin{equation} \label{E:difrate1}
\frac{\partial^2\Gamma}{\partial E_N \partial L_N} =
4\pi r^2 \int d^3\mathbf{v}~v_z \delta(E^\prime - E_N)
\delta(L^\prime - L_N) f(\mathbf{r},\mathbf{v})~,
\end{equation}
where the volume integral extends over the region $v_z > 0$ and
$E^\prime$ and $L^\prime$ are given by
\begin{subequations} \label{E:Ndef}
\begin{eqnarray}
E^\prime &=& \frac{1}{2} v^2~, \\
\label{E:EN}
L^\prime &=& |\mathbf{r} \times \mathbf{v}|~.
\label{E:LN}
\end{eqnarray}
\end{subequations}
We can use Eq.~(\ref{E:maxDF}) and the delta functions to evaluate
the integral to find
\begin{equation} \label{E:difrate2}
\frac{\partial^2\Gamma}{\partial E_N \partial L_N} =
\frac{(8\pi)^{1/2}nL_N}{\sigma^3} e^{-E_N/\sigma^2}~.
\end{equation}
If $\sigma \ll c$, orbits near the SBH will be insensitive to the
value of $E_N$ and we can integrate over this variable to yield a
differential rate
\begin{equation} \label{E:difrate3}
\frac{\partial \Gamma}{\partial L_N} = \int_{0}^{\infty}
\frac{\partial^2\Gamma}{\partial E_N \partial L_N} dE_N =
\frac{(8\pi)^{1/2}nL_N}{\sigma}~.
\end{equation}
The divergence of this rate as $\sigma \to 0$ results from
gravitational focusing, which would channel all stars into the SBH in
the absence of tangential velocities.

Before proceeding to the relativistic calculation, let us review the
Newtonian predictions.  Although the event horizon is fundamentally a
relativistic concept, the ``Newtonian'' prediction would be that a
star is directly captured by the SBH if the pericenter of its orbit is
less than the Schwarzschild radius (\ref{E:RS}).  The pericenter of a
parabolic ($E_N = 0$) orbit with specific angular momentum $L_N$ is
$L_{N}^2/2GM$.  Equating this to the Schwarzschild radius, a star is
directly captured if $L_N \leq L_{\rm cap} \equiv 2GM/c$, which
according to Eq.~(\ref{E:difrate3}) implies a capture rate
\begin{eqnarray}
\Gamma_{\rm cap} &=& \int_{0}^{L_{\rm cap}} \frac{\partial \Gamma}{\partial L_N}
dL_N = \frac{(32\pi)^{1/2}n(GM)^2}{\sigma c^2}
\nonumber \\
&=& 2.1 \times 10^{-6} {\rm yr}^{-1} \left( \frac{M}{10^6 M_\odot} \right)^2
\left( \frac{n}{10^5 {\rm pc}^{-3}} \right)
\nonumber \\
\label{E:RcapN}
&& \times \left( \frac{\sigma}{100~{\rm km/s}} \right)^{-1}
\end{eqnarray}
A star will be tidally disrupted if the pericenter of its orbit is
less than the tidal-disruption radius $r_{\rm TD}$ (\ref{E:TDconNR}),
which implies an angular momentum
\begin{equation} \label{E:LTD}
L_N \leq L_{\rm TD} \equiv \left(
\frac{(2M)^{4/3}GR_\ast}{m_{\ast}^{1/3}} \right)^{1/2}~.
\end{equation}
This implies a TDE rate
\begin{eqnarray}
\Gamma_{\rm TD}^\prime &=& \int_{0}^{L_{\rm TD}}
\frac{\partial \Gamma}{\partial L_N} dL_N =
\frac{(8\pi)^{1/2}nGMR_\ast}{\sigma} \left( \frac{2M}{m_\ast} \right)^{1/3}
\nonumber \\
&=& 6.3 \times 10^{-5} {\rm yr}^{-1} \left( \frac{M}{10^6 M_\odot} \right)^{4/3}
\left( \frac{n}{10^5 {\rm pc}^{-3}} \right)
\nonumber \\
\label{E:RTDN}
&& \times \left( \frac{\sigma}{100~{\rm km/s}} \right)^{-1}
\end{eqnarray}
This rate agrees with that in Eq.~(16b) of Frank and Rees
\cite{Frank:1976uy} which applies when the critical radius at which
the loss cone refills on a dynamical time exceeds the SBH's radius of
influence.  If TDEs can only be observed when the tidal debris is not directly
captured by the SBH, the observed TDE rate will be
$\Gamma_{\rm TD} = \Gamma_{\rm TD}^\prime - \Gamma_{\rm cap}$.  Since
$\Gamma_{\rm TD}^\prime \propto M^{4/3}$ while $\Gamma_{\rm cap}
\propto M^2$, the TDE rate will vanish for $M \geq M_{\rm max}$
(\ref{E:Mmax}) at which $r_{\rm TD} = r_s$.

We can use this same differential rate $\partial \Gamma/\partial L_N$
(\ref{E:difrate3}) to calculate the relativistic direct-capture and
TDE rates.  However, we must now rely on the Monte Carlo simulations
of Sec.~\ref{S:sims} to determine whether stars are directly captured
or tidally disrupted, instead of the simple Newtonian expressions for
$L_{\rm cap}$ and $L_{\rm TD}$ given above.  The simulated orbits have
a maximum angular momentum $L_{\rm max} \equiv (2GMr)^{1/2} \sin
\Theta_{\rm max}$.  The total rate at which stars on these orbits enter
a sphere of radius $r = 2000 M$ is
\begin{equation} \label{E:totrate}
\Gamma_{\rm tot} = \int_{0}^{L_{\rm max}} \frac{\partial \Gamma}{\partial L_N}
dL_N = \frac{(8\pi)^{1/2}nGMr}{\sigma} \sin^2 \Theta_{\rm max}~.
\end{equation}
The rate $\Gamma_{\rm cap}$ at which stars are directly captured by
the SBH is found by multiplying this total rate $\Gamma_{\rm tot}$ by
the fraction $F_{\rm cap}$ of simulated geodesics that cross the event
horizon.  The TDE rate $\Gamma_{\rm TD}$ is similarly found by
multiplying $\Gamma_{\rm tot}$ by the fraction $F_{\rm TD}$ of orbits
that violate the relativistic criterion (\ref{E:TDcon}) for tidal
disruption.  If $r$ and $\Theta_{\rm max}$ are chosen large enough,
these fractions $F
\propto (r\sin^2 \Theta_{\rm max})^{-1}$ so that the physical rates are
independent of our choice of initial conditions.

\begin{figure}[t!]
\begin{center}
\includegraphics[width=3.5in]{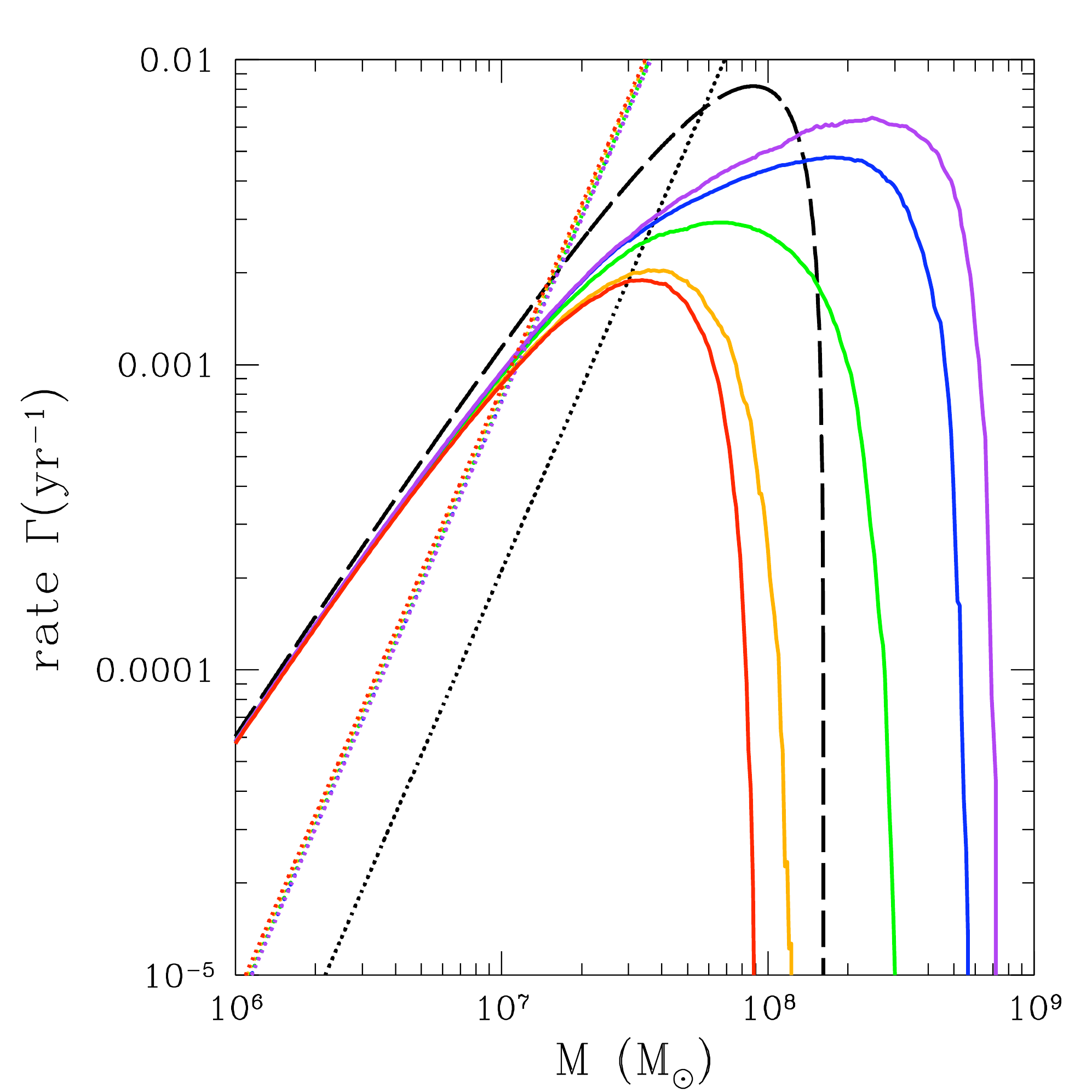}
\end{center}
\caption{The rates at which stars are directly captured (dotted lines)
and tidally disrupted (solid and dashed curves) by SBHs of mass $M$ in
constant-density cores with $n = 10^5~{\rm pc}^{-3}$ and $\sigma =
100$ km/s.  The black curves show the Newtonian rates of
Eqs.~(\ref{E:RcapN}) and (\ref{E:RTDN}), while the colored curves show
the relativistic rates for SBHs with spins $a/M = 0$ (red), 0.5
(orange), 0.9 (green), 0.99 (blue), and 0.999 (purple).  The capture
rates mildly decrease with SBH spin, while for $M \geq 10^8 M_\odot$
the TDE rates greatly increase with SBH spin.}
\label{F:Grate}
\end{figure}

In Fig.~\ref{F:Grate}, we show the direct-capture rate $\Gamma_{\rm
cap}$ and TDE rate $\Gamma_{\rm TD}$ as functions of SBH mass $M$ for
our fiducial choices $n = 10^5~{\rm pc}^{-3}$ and $\sigma = 100$ km/s.
The Newtonian prediction for $\Gamma_{\rm cap}$ underestimates the
true relativistic capture rate by about a factor of 4.  If we had used
the true specific angular momentum $L_z = 4GM/c$ for marginally bound
geodesics of a Schwarzschild SBH as the upper limit of the integral in
our Newtonian prediction (\ref{E:RcapN}), we could have nearly
reproduced the correct relativistic result.  The capture rate is
nearly independent of SBH spin as indicated by the colored dotted
lines lying almost on top of each other.  This is surprising, since
the specific angular momentum $L_z$ for prograde (retrograde)
marginally bound equatorial orbits varies from $4M$ ($-4M$) to $2M$
($-4.828M$) as $a/M$ increases from 0 to 1.  Near perfect cancellation
over orbital orientation ($\theta, \Theta, \Phi$) must occur for
the capture rate $\Gamma_{\rm cap}$ to be so mildly dependent on spin,
but we do not see any obvious reason for this to be the case.  Young {\it et al.}
\cite{Young:1977} calculated the ratio of the capture rate $\Gamma_{\rm cap}(a)$
for Kerr SBHs of spin $a$ to the capture rate $\Gamma_{\rm cap}(0)$ for nonspinning
SBHs.  Equation (B2) of their paper shows that this ratio is approximately given by
\begin{equation} \label{E:capratio}
\frac{\Gamma_{\rm cap}(a)}{\Gamma_{\rm cap}(0)} =  1 -
0.0820 \left( \frac{a}{M} \right)^2 + 0.0717 \left( \frac{a}{M} \right)^4
- 0.0864 \left( \frac{a}{M} \right)^6.
\end{equation}
The small numerical values of the coefficients in this expression indicate the weak
dependence of the direct-capture rate on SBH spin; the ratio is between 0.9 and unity
over the full range of spins $0 \leq a/M \leq 1$.

The TDE rate $\Gamma_{\rm TD}$ exhibits a much stronger dependence on
SBH spin, as illustrated by the strongly varying solid colored curves
in Fig.~\ref{F:Grate}.  At small SBH masses, $r_{\rm TD} \gg M$ and
the TDE rate for all spins converges to the Newtonian result as
expected.  However, as $M$ increases, tidal disruption occurs closer
to the SBH where the Newtonian approximation is increasingly invalid.
This is most glaringly apparent for masses $M \gtrsim M_{\rm max}$ of
Eq.~(\ref{E:Mmax}) for which tidal disruption would not be possible in
the Newtonian limit.  The true maximum mass, where the solid colored
curves in Fig.~\ref{F:Grate} intersect $\Gamma_{\rm TD} = 0$, is given
as a function of spin in Fig.~\ref{F:Mmax}.  Since $\Gamma_{\rm
TD}^\prime \propto M^{4/3}$, these massive SBHs are capable of tidally
disrupting even more stars than their less massive counterparts.
Although the spins $a/M = 0.99$ and $0.999$ depicted by the blue and
purple curves in Fig.~\ref{F:Grate} may seem extreme, the simple
scenario of growing a SBH from a standard thin accretion disk leads
to a limiting spin $a/m \simeq 0.998$ quite close to the purple curve
\cite{Thorne:1974ve}.  Although uncertain, cosmological predictions
for SBH spin distributions can also be peaked near these large values
\cite{Berti:2008af}.  The primary conclusions to draw from our analysis
are that relativistic corrections to the TDE rate can alter
predictions by a factor of several for $M \gtrsim 10^7 M_\odot$, and
that they can allow TDEs to occur for SBHs as large as $\sim 10^9 M_\odot$.

\subsection{Real galaxies} \label{SS:real}

Following Frank and Rees \cite{Frank:1976uy}, the rates we calculated
in the preceding subsection assumed that galaxies had constant-density
cores outside the SBH's radius of influence,
\begin{equation} \label{E:rh}
r_h \equiv \frac{GM}{\sigma^2} = 0.43~{\rm pc} \left(
\frac{M}{10^6 M_\odot} \right) \left( \frac{\sigma}{100~{\rm km/s}}
\right)^{-2} .
\end{equation}
Real galaxies with either power-law or core profiles have
mass-density profiles $\rho(r)$ that monotonically decrease with
radius.  This raises the question of what is the appropriate number
density $n$ to insert in our expressions for direct-capture and TDE
rates.  Frank and Rees \cite{Frank:1976uy} argued that the appropriate
density to use is that at the critical radius $r_{\rm crit}$ at which
stellar diffusion can refill the loss cone of tidally disrupted orbits
on a dynamical time.  A very crude estimate of this density can be
made by assuming that $r_{\rm crit} \simeq r_h$, an assumption roughly
true for real galaxies as indicated by Fig.~6 of Wang and Merritt
\cite{Wang:2003ny}.  If we further assume that the density profile of
galactic centers is that of a single isothermal sphere,
\begin{equation} \label{E:rhoISO}
\rho(r) = \frac{\sigma^2}{2\pi Gr^2}~,
\end{equation}
then inserting $n = \rho(r_h)/m_\ast$ into Eq.~(\ref{E:RTDN}) implies 
\begin{equation} \label{E:rhrcrit}
\Gamma_{\rm TD}^\prime \simeq 1.3 \times 10^{-3} {\rm yr}^{-1}
\left( \frac{M}{10^6 M_\odot} \right)^{-2/3}
\left( \frac{\sigma}{100~{\rm km/s}} \right)^5~.
\end{equation}
Wang and Merritt \cite{Wang:2003ny} use the isotropic distribution
function appropriate for a single isothermal sphere to calculate the
true rate at which the loss cone is refilled by stellar diffusion.
They find that their results are well approximated by the fit
\begin{equation} \label{E:WMrate}
\Gamma_{\rm TD}^\prime \simeq 2.5 \times 10^{-3} {\rm yr}^{-1}
\left( \frac{M}{10^6 M_\odot} \right)^{-1}
\left( \frac{\sigma}{100~{\rm km/s}} \right)^{7/2}~.
\end{equation}
If we combine this estimate with a recent determination of the
relation between SBH mass and host-galaxy velocity dispersion
\cite{Schulze:2010yy},
\begin{equation} \label{E:Msig}
\frac{M}{10^6 M_\odot} = 7.58 \left( \frac{\sigma}{100~{\rm km/s}}
\right)^{4.32},
\end{equation}
we arrive at a final TDE rate of
\begin{equation} \label{E:Frate}
\Gamma_{\rm TD}^\prime \simeq 4.8 \times 10^{-4} {\rm yr}^{-1}
\left( \frac{M}{10^6 M_\odot} \right)^{-0.19}
\end{equation}
in the Newtonian limit.  This estimate should be reasonable for the
power-law galaxies that dominate the total TDE rate; the core galaxies
that host the most massive SBHs have TDE rates $\Gamma_{\rm TD}^\prime
\simeq 10^{-5}~{\rm yr}^{-1}$ about an order of magnitude below that
of comparable-mass power-law galaxies \cite{Wang:2003ny}.

\begin{figure}[t!]
\begin{center}
\includegraphics[width=3.5in]{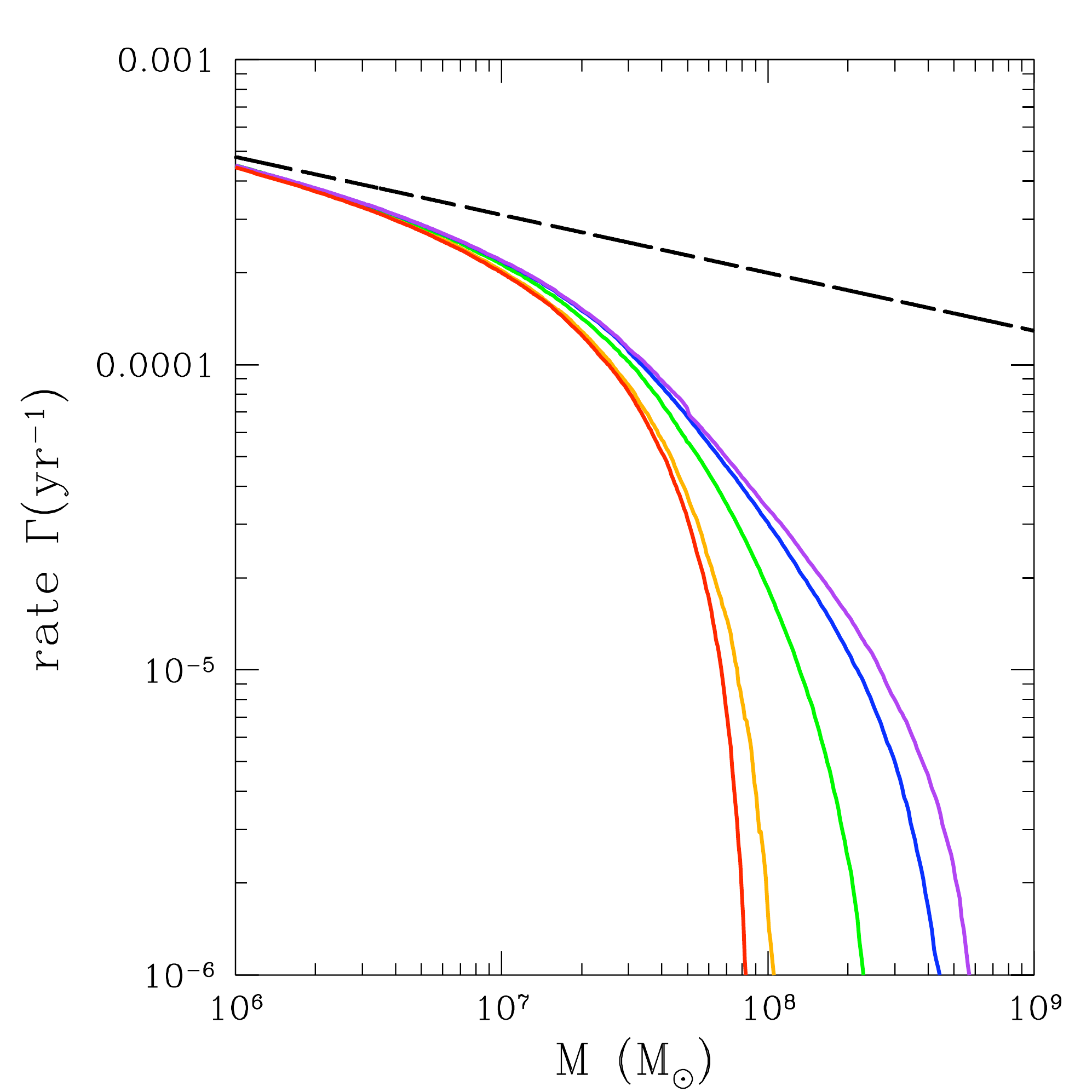}
\end{center}
\caption{The rates at which stars are tidally disrupted by SBHs of
mass $M$ in power-law galaxies obeying the $M-\sigma$ relation.  The
dashed black line is the prediction of Wang and Merritt
\cite{Wang:2003ny} for $\Gamma_{\rm TD}^\prime$ with an updated
$M-\sigma$ relation.  The colored curves show our relativistic
corrections $\Gamma_{\rm TD}$ to this prediction. The TDE rate increases
with SBH spin, with the given curves corresponding to
$a/M = 0$ (red), 0.5 (orange), 0.9 (green), 0.99 (blue), and 0.999
(purple).}
\label{F:Rrate}
\end{figure}

In Fig.~\ref{F:Rrate}, we show how the direct capture of stars by
spinning SBHs changes this prediction.  This figure was prepared with
the same set of Monte Carlo simulations described in
Sec.~\ref{S:sims}.  Although there are considerable differences
between the Newtonian predictions of Eqs.~(\ref{E:RTDN}) and
(\ref{E:Frate}), these differences result from different treatments of
the stellar populations far from the SBH.  We may therefore simply
renormalize our relativistic predictions $\Gamma_{\rm TD} = F_{\rm TD}
\Gamma_{\rm tot}$ of the previous subsection by dividing by
Eq.~(\ref{E:RTDN}) and multiplying by Eq.~(\ref{E:Frate}) at each SBH
mass $M$.  Direct capture reduces the predicted TDE rate by a factor
$\sim 2/3~(1/10)$ at $M = 10^7~(10^8) M_\odot$.  Although TDEs are
very rare for large SBH masses, they are still possible for $M <
M_{\rm max} \simeq 10^9 M_\odot$.  Since SBHs with masses $M \simeq
10^9 M_\odot$ predominantly live in galaxies with cored profiles,
Fig.~\ref{F:Rrate} may somewhat underestimate TDE rates at these
masses since the stellar density will not fall as steeply with $r$ as
the single isothermal profile of Eq.~(\ref{E:rhoISO}).

\section{Discussion}  \label{S:disc}

Astronomers have sought to observe the electromagnetic flares
associated with TDEs ever since this possibility was proposed by Rees
\cite{Rees:1988bf}.  Several potential TDEs were discovered over the
past 15 years by the Roentgen Satellite (ROSAT) \cite{Bade:1996} and the Galaxy Evolution
Explorer \cite{Gezari:2006fe}, and the recent discovery of
additional TDEs by both the Sloan Digital Sky Survey 
\cite{vanVelzen:2010jp} and Swift
\cite{Burrows:2011dn,Levan:2011yr,Bloom:2011xk,Cenko:2011ys} has
renewed interest in this phenomenon.  While individual TDEs may provide
new insights into SBH accretion physics, the large samples that may soon
be available \cite{vanVelzen:2010jp} will uniquely probe the whole
population of both active and quiescent SBHs.  While overall TDE rates
depend on stellar populations at galactic centers, the upper bound on
the mass $M$ of SBHs capable of tidal disruption is a sensitive
measure of SBH spins.  For $M \gtrsim 10^7 M_\odot$, tidal disruption
occurs deep enough in the SBHs potential well that Newtonian gravity
is no longer valid.  Furthermore, there is no reason to expect the
orbital angular momenta of tidally disrupted stars to align with SBH
spins.  For both these reasons, accurate calculations of TDE rates
require evaluation of the relativistic tidal tensor $C_{ij}$ on a
representative sample of generically oriented Kerr geodesics.

We have performed a series of Monte Carlo simulations that provide
this required sample.  We use this sample to calculate TDE rates for
spinning SBH as a function of their mass $M$, both in constant-density
cores and in isothermal spheres that approximate real power-law
galaxies.  We find that for $M \gtrsim 10^7 M_\odot$, a significant
fraction of stars will be directly captured by the SBH's event horizon
instead of being tidally disrupted and subsequently accreted.  This
will reduce the observed TDE rate assuming that directly captured
stellar debris will not have the chance to radiate appreciably before
being swallowed by the SBH.  Above $M \simeq 10^8 M_\odot$, only
highly spinning ($a/M \gtrsim 0.9$) SBHs will be able to produce
observable TDEs.  Theory \cite{Berti:2008af} and observation
\cite{Reynolds:1998ie,Brenneman:2006hw} suggest that most SBHs
may have such large spins, but further observations are needed to
investigate this possibility.  A future survey like the Large Synoptic Survey
Telescope \cite{:2009pq} that finds thousands of TDEs may provide
important constraints on the distribution of SBH spins.

{\bf Acknowledgements.} We would like to thank Mike Blanton, Glennys
Farrar, Andrei Gruzinov, David Merritt, Maryam Modjaz, Sterl Phinney,
and Scott Tremaine for useful conversations.

\end{document}